\documentclass[aps,twocolumn,groupedaddress,superscriptaddress,floatfix,amsmath,amssymb,prb]{revtex4-1}
\usepackage{amsmath}
\usepackage{amssymb}
\usepackage{graphicx}
\usepackage{textcomp}
\usepackage{bm}

\usepackage{booktabs}
\usepackage{siunitx}

\usepackage[usenames,dvipsnames]{color}	

\begin{document}

\title{Electromechanical Quantum Simulators}
    
\author{F. Tacchino$^{+}$}
\affiliation{Dipartimento di Fisica, Universit\`a di Pavia, via Bassi 6, I-27100 Pavia, Italy}
\author{A. Chiesa$^{+}$}
\affiliation{Dipartimento di Scienze Matematiche, Fisiche e Informatiche, Universit\`a di Parma, I-43124 Parma, Italy}
\affiliation{Institute for Advanced Simulations, Forschungszentrum J\"ulich, 52425 J\"ulich, Germany}
\author{M. D. LaHaye}
\affiliation{Department of Physics, Syracuse University, Syracuse NY 13244-1130, USA}
\author{S. Carretta}
\email{stefano.carretta@unipr.it}
\affiliation{Dipartimento di Scienze Matematiche, Fisiche e Informatiche, Universit\`a di Parma, I-43124 Parma, Italy}
\author{D. Gerace}
\email{dario.gerace@unipv.it}    
\affiliation{Dipartimento di Fisica, Universit\`a di Pavia, via Bassi 6, I-27100 Pavia, Italy}

\begin{abstract} 
Digital quantum simulators are among the most appealing applications of a quantum computer. 
Here we propose a universal, scalable, and integrated quantum computing platform based on tunable 
nonlinear electromechanical nano-oscillators.
It is shown that very high operational fidelities for single and two qubits gates can be achieved in a minimal 
architecture, where qubits are encoded in the anharmonic vibrational modes of mechanical nanoresonators, 
whose effective coupling is mediated by virtual fluctuations of an intermediate superconducting artificial atom.
An effective scheme to induce large single-phonon nonlinearities in nano-electromechanical devices is explicitly discussed, thus opening
the route to experimental investigation in this direction.
Finally, we explicitly show the very high fidelities that can be reached for the digital quantum simulation of model Hamiltonians, 
by using realistic experimental parameters in state-of-the art devices, and considering the transverse field Ising model as a paradigmatic example. 
\end{abstract}

\maketitle

\section{Introduction}

{Quantum simulators are being pursued in a number of different systems, ranging from cold atoms to  trapped ions, impurities in semiconductors or superconducting circuits \cite{Georgescu2014}. Among the various platforms, unprecedented progress is currently ongoing towards achieving a scalable architecture for quantum information processing that is based on purely superconducting qubits \cite{Clarke2008,Mariantoni2011,Devoret2013,Gambetta2017}.
These developments are particularly relevant in view of practically realizing \textit{digital} quantum simulators, i.e. systems able to simulate the dynamical evolution of any model that can be represented as a sum of local Hamiltonian terms \cite{Lloyd1996}. 
However,  scalability of a digital quantum processor beyond a few qubits register requires gate fidelities that are still incompatible with the relatively short coherence times, even in state-of-the-art superconducting elements \cite{Koch2007,Rigetti2012}. Hence, several proposals have been putting forward the idea of developing hybrid quantum circuits, in which superconducting elements are efficiently and coherently coupled to other degrees of freedom with possibly longer coherence times \cite{Xiang2013,Carretta2013,Chiesa2014}. Recently, it has been suggested that hybrid optomechanical devices could also be used for quantum information processing \cite{Rips2013}. 

Here we envision a novel hybrid architecture to efficiently implement a digital quantum simulator, which is based on electromechanical elements coupled to superconducting circuits. At difference with the existing literature, we propose to use anharmonic and tunable nanomechanical resonators (NRs) to encode the quantum information, while virtual fluctuations of superconducting elements such as transmons are only employed to perform two-qubit gates, making their $T_2$ time essentially irrelevant. The proposal is motivated by the recent progress in hybrid quantum electromechanical systems realized in superconducting circuits, with a focus on investigating transmon-nanoresonator interactions \cite{Lahaye2009,OConnel2010,Sillanpaa2013,Sillanpaa2015,Rouxinol2016,Chu2017}. As model nano-electromechanical oscillators we consider either nanomembranes \cite{Teufel2011,Palomaki2013} and nanotubes \cite{Schneider2014,Moser2014}, or graphene sheets \cite{Singh2014,Weber2014,Song2014}, which have been shown to display remarkably low damping and decoherence rates. Large tunability of their resonance frequencies \cite{Singh2014,Weber2014,Postma2005,Kozinsky2006} as well as their nonlinear properties \cite{Kozinsky2006,Agarwal2006} have been experimentally shown. In addition to the existing proposals to considerably increase the anharmonic behavior of mechanical resonators \cite{Khan2013,Rips2014}, we hereby test an efficient scheme to induce a very large single-phonon shift of the fundamental vibrational frequency by dispersive coupling to a superconducting element \cite{Nori2010}. 

As a direct comparison with the currently dominant technology based on superconducting qubits, the electromechanical qubit encoding would not only allow achieving ultra-high gate fidelities in excess of $99.9\%$, as quantitatively shown in the following with state-of-art parameters, but also $T_2 / T_{\mathrm{gate}} > 10^4$, where $T_{\mathrm{gate}}$ represents the average single- and two-qubits overall gating time. The latter is a key figure of merit in view of scalability of any proposed platform for the realization of digital quantum simulation in which a long sequence of concatenated operations is required, which represents a sensitive improvement over previous proposals \cite{LasHeras2014,Mezzacapo2014,Chiesa2015} and state-of-art realizations \cite{Barends2014,Abdi2015,Barends2015,Salathe2015} in superconducting platforms.
}

\section{Hybrid electromechanical platform} 

\begin{figure}[t]
\centering
\includegraphics[width=0.46\textwidth]{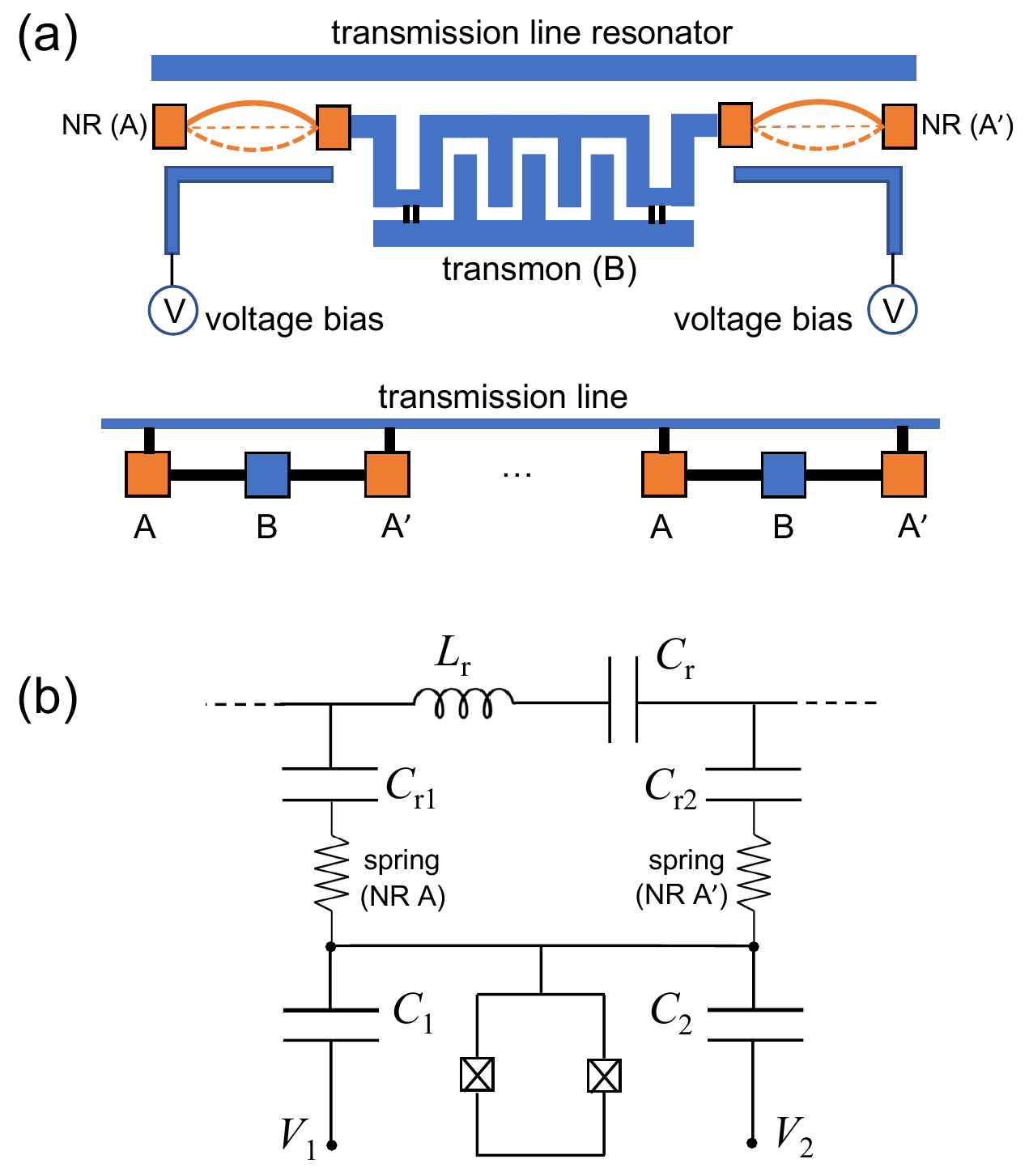} 
\caption{(a) Schematic representation of the circuit building block for digital quantum simulation proposed in this work: two electromechanical oscillators (NR) mutually coupled through a transmon within a superconducting circuit; each element is tunable through either external voltage bias (the NR) or external magnetic fields (the transmon), {and ground state cooling of the NRs can be achieved by coupling to a transmission line resonator, or a lumped LC circuit element}; the intrinsic scalability of this set-up is sketched below. (b) Analog circuit model of the elementary building block shown in (a).}
\label{fig:scheme}
\end{figure}

The fundamental unit of our architecture is given by a pair of electromechanical NRs mutually coupled to a nonlinear circuit element, here assumed to be a transmon. A schematic representation of this elementary building block as well as the corresponding circuit model are shown in Fig.\ \ref{fig:scheme}. Notice the straightforward scalability of this set-up according to a sequential repetition of the building block. A superconducting resonator or a lumped element (schematically illustrated in the Figure) can be taken into account as a further element to be weakly coupled to the NR for ground state cooling \cite{Teufel2011} (i.e., qubit initialization), while the transmon could also be employed for read out of each qubit state. In the following calculations, we will assume the NRs to be cooled to their ground states, without explicitly including the LC circuit in the model.

\subsection{Basic model}

 The elementary unit of the electromechanical platform can then be modeled through a second-quantized Hamiltonian, where the free mechanical resonators and the isolated transmon are described by ($\hbar = 1$)
\begin{equation}
H_{0}= \sum_i \left[ \omega_i b^\dagger_i b_i + H_{nl,i} \right] + \frac{\Omega} {2}\sigma_z \, ,
\label{eq:H0}
\end{equation}
with $b_i$ ($b_i^\dagger$) representing bosonic annihilation (creation) operators, and $H_{nl,i}$ giving the necessary anharmonicity to isolate the two lowest energy levels of the NRs, where the qubits are encoded. 
In order to keep our analysis simple, we will only require a shift of the lowest lying Fock energy levels, i.e. a diagonal term on the Fock basis, modeled by $H_{nl,i} = \beta b^\dagger_i b^\dagger_i b_i b_i$  (see App.~\ref{app:NonLin} for details).
The last term in $H_{0}$ describes the transmon as a pure quantum two-level system \cite{Koch2007}, $\sigma_\alpha$ ($\alpha=x,y,z$) representing the Pauli matrix. 
The interaction between mechanical oscillators and the transmon is modeled as \cite{Sillanpaa2013,coupling}
\begin{equation}
H_{int} = \sum_{i=1}^2 g_{i} \left(b_i+b_i^\dagger\right)\sigma_x \, . 
\label{eq:Hint}
\end{equation}
In the following, the electromechanical resonators frequencies will be set below 100 MHz, while the transmon frequency in the 2-10 GHz range. Notice that such an energy mismatch does not allow to employ the rotating wave approximation in $H_{int}$. 
Moreover, all the transmon excitations will only appear as virtual ones, while working with low-occupancy bosonic states. By a realistic choice of $\beta$, $g_i$ and $\Delta_i = \omega_i - \Omega$, the dynamics is effectively restricted to the computational basis that we will be considering. Dissipation and pure dephasing effects are fully included in our model by solving for the density matrix master equation
\begin{equation}
\partial_t \rho = \, i [ {\rho}, \hat{H}_{tot}(t)] + \mathcal{L}_{TR}[\rho] + \sum_i \mathcal{L}_i [\rho] \, ,
\label{eq:mastereq}
\end{equation}
with Lindblad terms acting individually on the electromechanical NRs, i.e. 
$\mathcal{L}_i[\rho] = \gamma_{i} \mathcal{D}(b_i)[\rho] + \gamma_{i,d} \mathcal{D}(b_i^\dagger b_i)[\rho]$,
and on the transmon, i.e. 
$\mathcal{L}_{TR}[\rho] = \gamma_{TR} \mathcal{D}(\sigma_-)[\rho] + \gamma_{TR,d} \mathcal{D}(\sigma_z)[\rho]$, 
respectively, where $\mathcal{D}(a)[\rho] = a\rho a^\dagger - 0.5 \{ a^\dagger a, \rho\}$. 

\subsection{Mechanical anharmonicity} 
\label{sec:mech_anharm}

The required degree of anharmonicity for the vibrating oscillators to be defined as qubits deserves a separate discussion. On one hand, anharmonic contributions to the mechanical vibrational eigenstates can be experimentally implemented through the use of intrinsic mechanical nonlinearities \cite{Postma2005,Kozinsky2006}, or by using static external electric fields~\cite{Kozinsky2006,Agarwal2006,Khan2013,Rips2014}.
On the other hand, the required anharmonic shift to achieve a reliable qubit behavior amounts to about 1 MHz at least (see, e.g., the effect of this value on the gate fidelities, reported in App.~\ref{app:NonLin}). Since it is not clear if a regime of single-phonon nonlinearity can be achieved with the above mentioned tools, we hereby explore an alternative scheme based on dispersive coupling of an additional low-frequency superconducting element to each NR, without degrading the mechanical resonator's remarkable coherence properties. Notice that such an additional component would not be involved in mediating the interaction between NRs, as the transmon qubit in the schematic picture of Fig.~\ref{fig:scheme}a. 

\begin{figure} 
\centering
\includegraphics[scale=0.44]{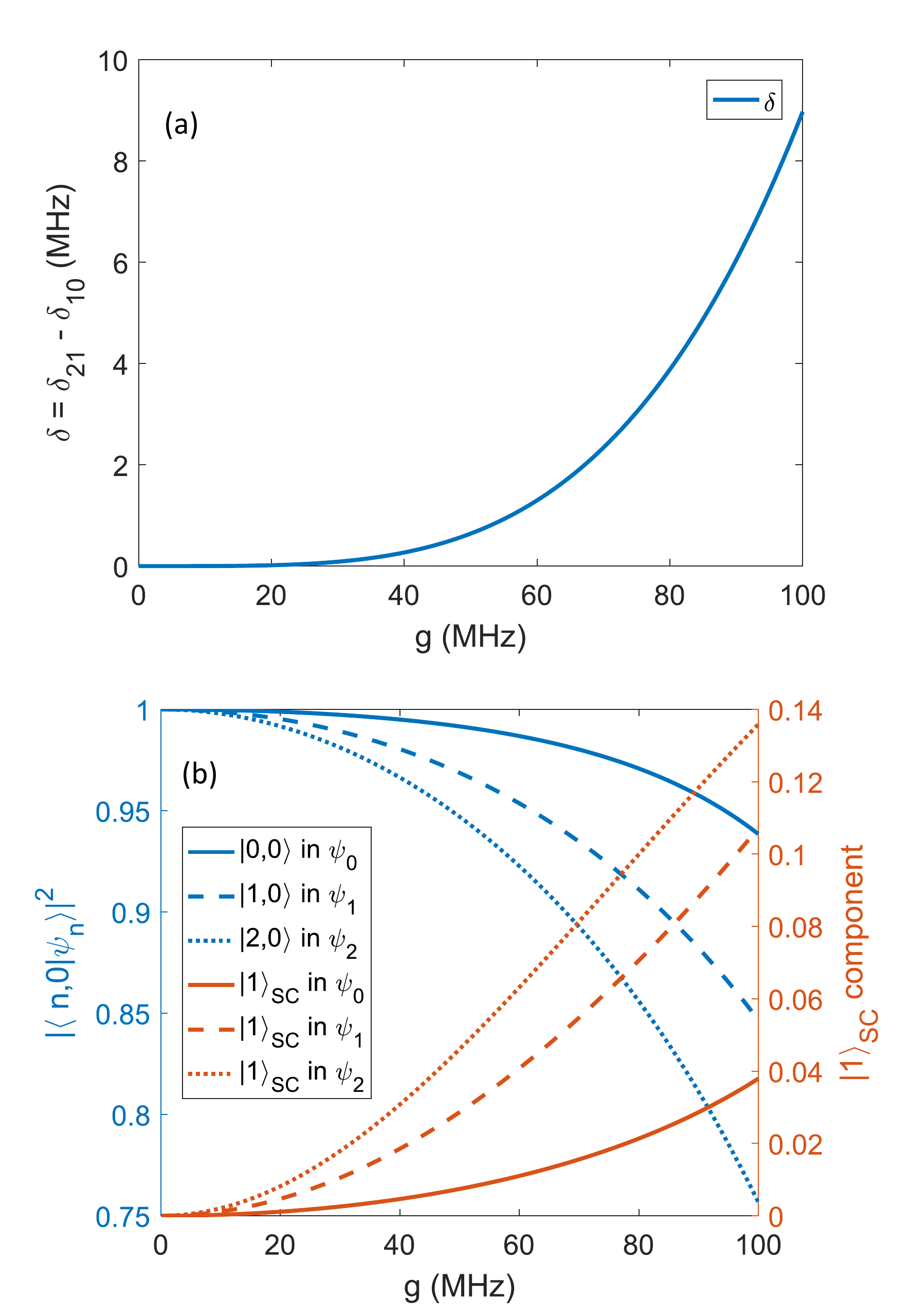} 
\caption{
Nonlinearity induced on a nanoresonator by a superconducting circuit. 
(a) Nonlinear shift between the ground-to-first and the first-to-second excited states transitions, $\delta = \omega_{21}-\omega_{10}$. 
(b) Components of the corresponding wave function. In both panels, $\omega_{NR} = 100\,$MHz, $\Omega_{SC}  = 500\,$MHz.}
\label{fig:NLfromSC}
\end{figure}

The key idea relies on engineering a nonlinear spectrum of the collective system composed by a NR and a superconducting (SC) element, such as a fluxonium, 
whose anharmonic energy levels would then be used for the definition of the physical qubit. Notice that this cannot be considered as a form of hybrid encoding: indeed, given the dispersive nature of the coupling, the two subsystems would not be treated on equal footing, with the NR degrees of freedom maintaining a predominant role. This is of fundamental importance when considering the possible effects of the SC circuit noise mechanisms on the qubit dissipation and coherence times (see App.~\ref{app:Single-Ph-Nonlin} for an extensive analysis). The coupled NR-SC element system is effectively described by a Rabi-like interaction term, with a Hamiltonian reading
\begin{equation}
H = \omega_{NR} b^\dagger b + \frac{\Omega_{SC}}{2}\sigma_z + g(b+b^\dagger)(\sigma_- + \sigma_+) \, .
\label{eq:H_NRSC}
\end{equation}
Notice that, due to the large detuning, $\Delta = \Omega_{SC}-\omega_{NR}$, and the large coupling strength, no rotating wave approximation can be made, as done, e.g., in previous works with the same spirit\cite{Nori2010}. It is worth stressing, once again, that the SC component here envisioned would not play the role of a mediator between different NRs, as the transmon element explicitly introduced in the fundamental building block of our set-up (see Fig.~\ref{fig:scheme}). Instead, it would be an additional element attached to each individual electromechanical oscillator, only used to slightly modify the structure of its excitation spectrum. As we are about to see, this difference manifests itself also in the parameters range required to obtain a significant nonlinear effect, which are very different from the ones at which a transmon qubit is typically operated. In fact, a SC resonance frequency, $\Omega_{SC}$, in the few hundreds of MHz range is required here, as well as a NR-SC element coupling rate, $g$, in the few tens of MHz, which is basically opposite to the large $\Omega$ ($1-10$ GHz range) and small $g$ (a few MHz) required for the transmon mediator. A nonlinear SC circuit (e.g., fluxonium) with low frequency and rather good dissipation and coherence properties, as compared to typical SC performances, has already been demonstrated experimentally \cite{Pop2014}. We assume that large NR-SC coupling strengths can be obtained capacitively by applying a sufficiently strong voltage bias. As a first step, we show in Fig.~\ref{fig:NLfromSC}a, the nonlinear shift, $\delta$, obtained as a function of $g$ by numerically diagonalizing the Hamiltonian in Eq.~\eqref{eq:H_NRSC}. As typical parameters for a proof of concept demonstration we assuemed $\omega_{NR}  = 100\,$MHz, $\Omega_{SC} = 500\,$MHz, and a suitable number of bosonic excitations to obtain numerical convergence for the eigenvalues and eigenstates of interest. The target value, from the results of Fig.~\ref{fig:RxVsU} in App.~\ref{app:NonLin}, would be $\delta \simeq 1\,$MHz, which is obtained for $g  \simeq50\,$MHz. While this value is rather large, it should be noted that $\sim 10$ MHz coupling rates between a transmon and a 70 MHz mechanical NR have already been demonstrated experimentally in Ref.~\onlinecite{Sillanpaa2013} {(see, in particular, the Methods summary in the quoted reference)}. In addition, the authors of the latter work discuss how it should be possible to improve the coupling strength to values larger than 25 MHz by {suitably changing} the circuit geometry. We can further add that going to larger charging energy, e.g.\ by using a fluxonium {qubit}, should be possible without decreasing the coherence time. All in all, increasing the charging energy of the superconducting element, in addition to improvements to geometry, should ultimately lead to couplings in the order of $50\,$MHz without too much effort. \\
For comparison with the numerical results, we also report hereby the analytical expression of the nonlinear shift as obtained to 4th order perturbation theory in the SC-NR oscillators coupling:
\begin{widetext}
\begin{equation}
\delta \simeq 2g^4 \left[\frac{2}{(\Omega-\omega)^2 (\Omega + \omega)} + \frac{2}{(\Omega+\omega)^2 (\Omega - \omega)} + \frac{1}{(\Omega+\omega)^3} + \frac{1}{(\Omega-\omega)^3} \right] \, ,
\label{eq:delta_analytic}
\end{equation}
\end{widetext}
in which different terms arise from the combined effect of rotating and counter-rotating terms in Eq.~\eqref{eq:H_NRSC}. In the rotating-wave approximation, all of the terms but the last one would cancel out.

To better understand the actual character of the collective excitations in the first three energy levels, $\psi_i$ (i.e.\ the designated computational basis plus one extra level), we report in Fig.~\ref{fig:NLfromSC}b the amplitude probability of the bare Fock eigenstates of the uncoupled harmonic oscillator on the corresponding new eigenstate of the coupled system, $p_n = |\langle n,0 |\psi_n\rangle|^2$. Here the second index in the bra vector indicates the SC in its ground state, $|0\rangle$. Moreover, on the right axis of the same Fig.~\ref{fig:NLfromSC}b we show the total probability, for each collective eigenstate, to find the SC in the $|1\rangle$ excited state, regardless of the state of the NR: this information gives an estimate of the amount of wavefunction leaking on the SC as a consequence of the coupling, that is the magnitude of the mixing of the bare degrees of freedom. As it can be seen, in the region of interest such mixing never exceeds $\sim 5\%$ for the relevant states in the computational basis, which would define the mechanical qubit. This is a relevant result, also in light of the analysis reported in App.~\ref{app:Single-Ph-Nonlin}, which guarantees that the NR performances  in terms of coherence are not significantly affected by the presence of the additional SC element.

\subsection{Effective qubit-qubit interaction}

From the last paragraph, we will hereby assume that the NR can be effectively considered as anharmonic mechanical oscillators with a single-phonon nonlinear shift in the few MHz range. Hence, an effective interaction Hamiltonian between the two electromechanical qubits can be derived by resorting to second order perturbation theory from the original model Hamiltonian, Eq. (\ref{eq:H0}), and by restricting to the portion of the total Hilbert space in which the transmon is in its ground state (details are provided in App.~\ref{app:Heff}), which describes the relevant dynamics of the pair of qubits restricted to the computational basis $\{|00\rangle, |10\rangle,|01\rangle, |11\rangle\}$:
\begin{equation}
H_{eff} = \sum_{i=1}^2 \left(\frac{\lambda_i}{2}\sigma^i_{z}\right) + \frac{\Gamma}{8} \left( \sigma^1_x\sigma^2_x +\sigma^1_y\sigma^2_y \right) + \text{const.}
\label{eq:Heff_main}
\end{equation}
where the $\sigma^i$ are Pauli operators in the computational basis of the NRs, $\lambda_i$ are single-qubit energy shifts (i.e., transmon-induced frequency renormalizations) and the effective XY coupling constant reads
\begin{equation}
\Gamma = \frac{4 g_1 g_2 \Omega(\omega_1^2 + \omega_2^2 -2\Omega^2)}{(\Omega^2 - \omega_1^2)(\Omega^2 - \omega_2^2)} \, . 
\end{equation}
We will use this $H_{eff}$ as the reference model to understand the behavior of the real system, for which we numerically solve the full master equation above.

\section{Single and two-qubit gates}

One of the key ingredients to perform single and two-qubit gates is the dynamical tuning of $\omega_1$ and $\omega_2$. This can be achieved by using external static and modulated electric fields, i.e. electrostatic potential energies $V$, which can locally act on a single NR as already shown experimentally \cite{Postma2005,Kozinsky2006,Weber2014,Singh2014}. 
In the idle configuration, the two NRs are significantly detuned: $|\omega_1-\omega_2| \gg \Gamma$, thus switching off the interaction term appearing in Eq. (\ref{eq:Heff_main}). Hence, the two qubits are decoupled and independent rotations of each of them can be implemented. The use of a high-frequency transmon helps improving the two qubits decoupling. \\
In particular \cite{Rips2013}, single qubit $R^i_z$-rotations can be performed by shifting the NRs oscillation frequency for the amount of time required to add the desired phase to the $n_i=1$ component of the wave-function. 
Other single-qubit rotations are obtained by an oscillating transverse field keeping a definite phase relationship with the quantum mechanical evolution of the system. Indeed, by choosing $H^{xy}_i(t) = V^{xy} (t) (b_i + b_i^\dagger)$, with $V^{xy} (t) = \Theta_{t_0} (\delta t)V^{xy}_0 \cos (\omega_i t + \theta)$ one can achieve either $R_x^i$ ($\theta = 0$) or $R_y^i$ ($\theta = \pi/2$) rotations. The total rotation angle equals the area under the pulse modulating the oscillation, i.e. $\phi_{xy} = -\int dt  V^{xy}_0\Theta_{t_0}(\delta t)$, where $\Theta_{t_0}({\delta t})$ is a step function of duration $\delta t$, starting at $t_0$. \\ 
Our setup can also straightforwardly implement the two-qubit entangling gate known as $\sqrt{\text{iSWAP}}$, described by the truth-table 
$|0 0 \rangle \to | 0 0 \rangle$ and 
$|1 1\rangle \to | 1 1 \rangle$, while 
$|1 0\rangle \to (| 1 0 \rangle + i | 0 1 \rangle)/\sqrt{2}$ and 
$| 0 1 \rangle \to (i | 1 0 \rangle +  | 0 1 \rangle)/\sqrt{2}$.
This gate is obtained by tuning the qubits to resonance, $\omega_i \rightarrow \omega_i' = \omega_i + \xi_i$ such that $\omega_1' = \omega_2'$, thus activating the effective interaction term in Eq.~\eqref{eq:Heff_main}. 
Indeed, the dynamics induced by the XY interaction in Eq.~\eqref{eq:Heff_main} corresponds to a $\sqrt{\text{iSWAP}}$ for a proper choice of the interaction time, $T_{gate} = \pi/|\Gamma|$, as we show in App.~\ref{app:Heff}.
The transmon-mediated interaction should then be turned off by bringing back the two NRs to their original frequencies. 
Together with single-qubit rotations, this constitutes a universal set of quantum operations. 
A further tuning knob of the present set-up is the dynamical variation of the transmon frequency induced by an external magnetic flux, which allows to considerably shorten the $\sqrt{\text{iSWAP}}$ gating time (see App.~\ref{app:Gates} for details).

\begin{figure}[t]
\includegraphics[width=0.44\textwidth]{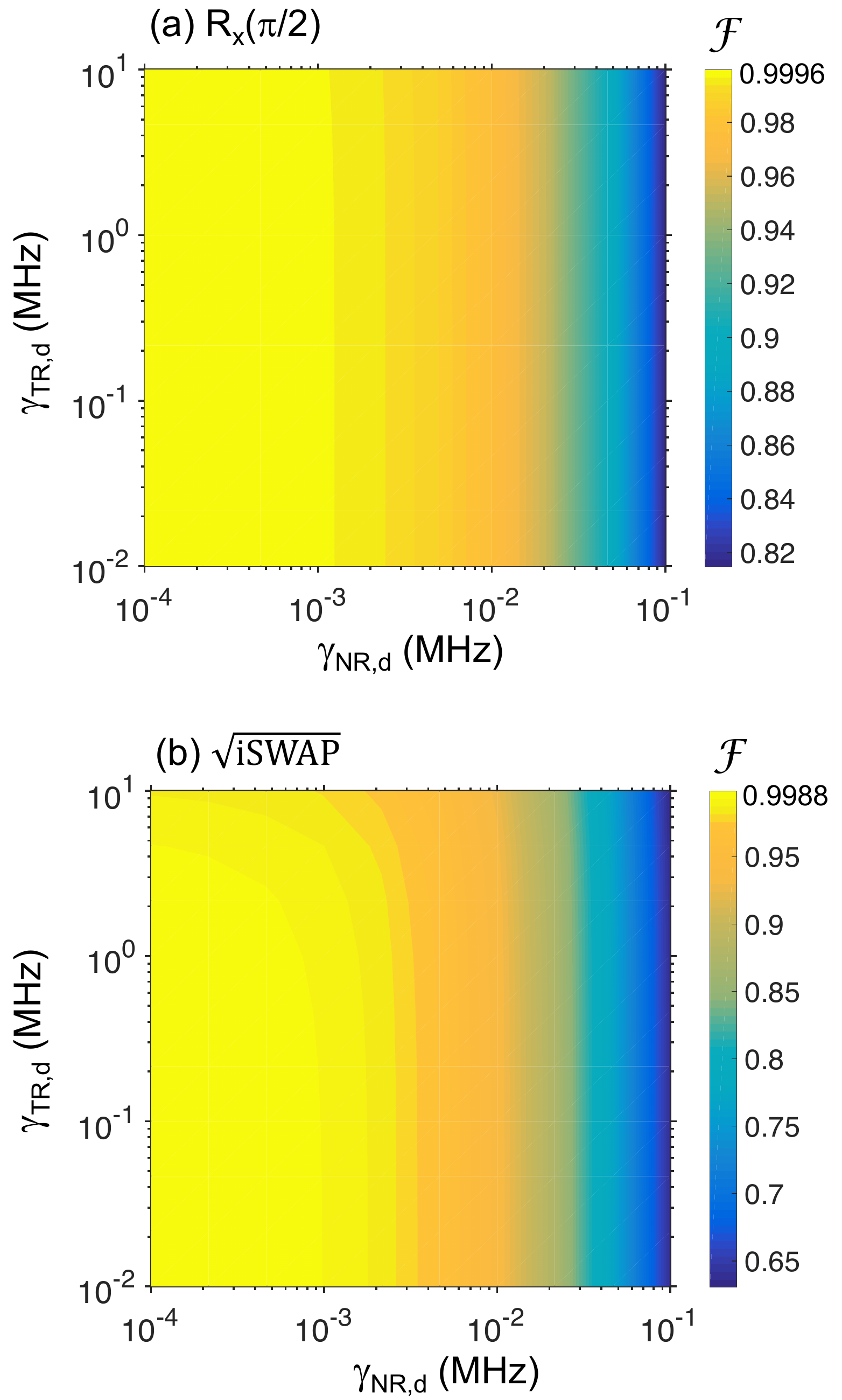} 
\caption{(a) Fidelity of a single qubit $x$-rotation by a $\pi/2$ angle, and (b) fidelity of the two-qubit $\sqrt{\text{iSWAP}}$ gate, as functions of the pure dephasing rates of the electromechanical resonators and the transmon, respectively. 
For the two-qubits gate, we assume $\gamma_{1,d} = \gamma_{2,d} =\gamma_{NR,d}$; all the other simulation parameters are reported in the text.}
\label{fig:FidScan}
\end{figure}

The density matrix master equation describing the full system is written as
$\partial_t \rho = \, i [ {\rho}, \hat{H}_{tot}(t)] + \mathcal{L}_{TR}[\rho] + \sum_i \mathcal{L}_i [\rho] \,$,
where $\hat{H}_{tot}(t) = \hat{H}_{0}+H_{int} + \hat{H}(t)$, and $H(t)$ includes all the time-dependent frequency shifts that are necessary to implement the gates. 
This master equation is numerically integrated, in the interaction picture, by using a standard Runge-Kutta algorithm.
A few representative examples of simulated single and two-qubit gates, together with a description of the required external pulses, are shown in App.~\ref{app:Gates}.  
Here we report the computed fidelities of illustrative single- and two-qubit operations in the presence of the main dissipation parameters of the model. 
The fidelity of a given gate is defined as $\mathcal{F}=\sqrt{\langle \psi | \rho | \psi\rangle}$ \cite{Nielsen}, where $\psi$ is the ideal target state and $\rho$ is the density matrix evolved through the full master equation. 
Realistic parameters for the hybrid circuit are assumed in all the simulations, such as: $\omega_1/ 2\pi = 85$ MHz and $\omega_2 /2\pi = 75$ MHz (idle configuration), $\beta /2\pi = 3$ MHz, $\Omega/2\pi$ is tuned from $10$ GHz (in idle configuration) down to $2.5$ GHz (when performing two-qubit gates), $g_1=g_2 =2\pi\cdot 6$ MHz, $\gamma_{1}=\gamma_2 = 2\pi\cdot 50$ Hz, $\gamma_{TR}/2\pi=$100 kHz. We also tested the effects of non-negligible thermal occupation of the nanomechanical modes (i.e., non perfect ground state cooling), showing that is does not appreciably affect the gate performances with our protocol (see App.~\ref{app:Thermal}). 
Finally, we checked that bosonic occupancies $n_i>1$ do not occur throughout the whole gate dynamics. \\
As benchmark operations, we show in Fig.~\ref{fig:FidScan} the calculated fidelities of a single qubit $R_x^1$-rotation, with $\phi_{x} = \pi/2$, and of the $\sqrt{\text{iSWAP}}$ gate, respectively, as a function of the pure dephasing rate (i.e., the reciprocal of the coherence time) for both the NRs and the transmon \cite{random}. We immediately notice a very weak dependence of $\mathcal{F}$ on $\gamma_{TR,d}$ and, as it could be expected, a more sensitive dependence on $\gamma_{NR,d}$. In particular, it is worth reminding that a value of $\gamma_{NR,d} / 2\pi\simeq 100$ kHz is utterly pessimistic for most of the electromechanical NRs, in particular nanomembranes and nanotubes, where total linewidths rather in the $0.1 - 1$ kHz range have been experimentally shown \cite{Singh2014,Weber2014,Moser2014}.   
The most remarkable and clear-cut message of these results is that, as expected from the virtual nature of transmon excitations, our scheme is intrinsically robust against transmon decoherence. Indeed, the results look practically insensitive to an increase of more than two orders of magnitude in $\gamma_{TR,d}$ from the most optimistic but still realistic \cite{Rigetti2012} value (i.e., $\gamma_{TR,d}/2\pi =10$ kHz, corresponding to a transmon $T_{2}$ time of 100 $\mu$s), for both single- and two-qubit gates. \\
{ Finally, we emphasize the comparison with transmon based technology: two-qubits gating times are currently on the order of 40-50 ns in state-of-art devices  when fidelities beyond 99\% are required \cite{Barends2014}, which means $T_2 / T_{\mathrm{gate}} \sim 10^3$ with current qubit coherence time \cite{Rigetti2012}. With our set-up, we have obtained $\mathcal{F}> 99\%$ with $T_{\mathrm{gate}} = 500$ (300) ns for single- (two-) qubit gates. These numbers show the potential impact of the proposed electromechanical platform to achieve $T_2 / T_{\mathrm{gate}} > 10^4$ for $T_{2,NR}\sim 10$ ms,  which is extremely promising and potentially better than state-of-the art transmon qubits.}



\begin{figure}[t]
\centering
\includegraphics[width=0.38\textwidth]{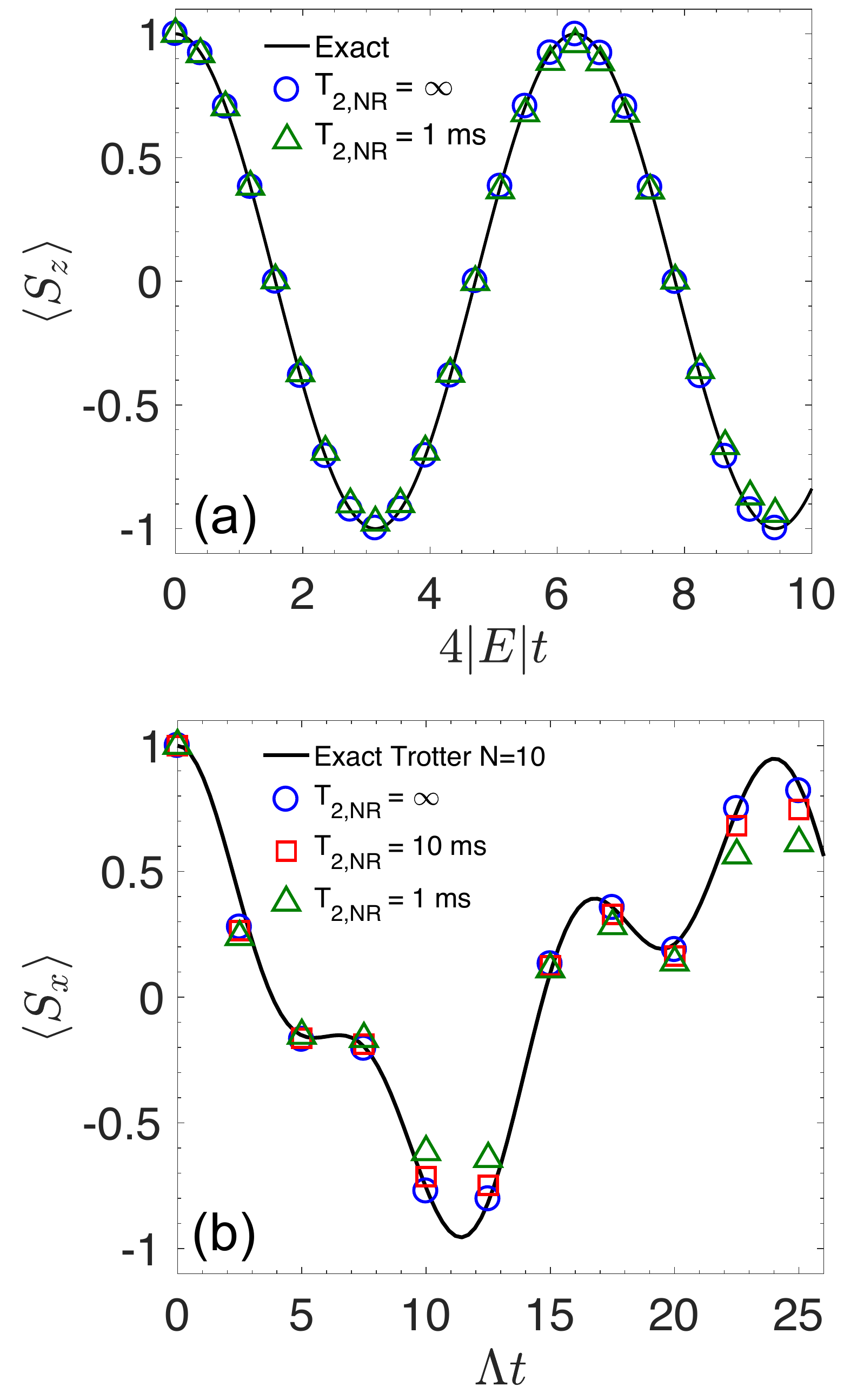} 
\caption{(a) Exact time evolution of the total magnetization of a $S=1$ system (full line) undergoing oscillations between opposite polarized states 
(describing, e.g., quantum tunneling across an anisotropy barrier), compared to the digital quantum simulation of the target Hamiltonian (\ref{eq:HS1}) 
for infinite and finite $T_2$ time of the electromechanical qubits, respectively (points). 
(b) Exact Trotter evolution ($N=10$) of the total $x$-polarization in the model Hamiltonian (\ref{eq:H_TIM}) (full line), compared to the corresponding 
quantum simulation for different values of the qubits $T_2$ time (points). 
}
\label{fig:simulator}
\end{figure}

\section{Digital quantum simulations}

The remarkable theoretical fidelities shown in the last Section for the elementary single and two-qubit gates are a crucial requirement for scaling up the quantum computation, e.g. to build a quantum simulator involving a long sequence of concatenated gates \cite{Georgescu2014}. 
Here we test the performances of a realistic proof-of-principle digital quantum simulation of illustrative models mapped onto spin-type Hamiltonians. This is done in analogy to previous works \cite{Santini2011,LasHeras2014,Chiesa2015}, by decomposing the time-evolution operator up to the instant $t$ into the product of $N$ terms, each evolving for short time intervals $\tau=t/N$, also named Trotter steps \cite{Lloyd1996}. 
For sufficiently small $\tau$, different terms of the target Hamiltonian commute, thus allowing us to decompose the target evolution into a sequence of quantum gates. In fact, we hereby focus on spin-type Hamiltonians, since most models of physical interest can be mapped onto a combination of local operators only involving one $\mathcal{H}_\alpha^{(1)}$ and two-body $\mathcal{H}_\alpha^{(2)}$ spin-terms, and the time evolution of these terms can be efficiently simulated through a proper sequence of one- and two-qubit gates. Indeed, the time evolution induced by $\mathcal{H}_\alpha^{(1)} \propto \sigma_\alpha^i$ directly corresponds to single-qubit rotations $R_\alpha^i$. Conversely, two-body terms of the form $\mathcal{H}_\alpha^{(2)} \propto \sigma_\alpha^1 \sigma_\beta^2$ can be obtained by combining the XY evolution given by the second term in Eq. (\ref{eq:Heff_main}) with single qubit rotations (as explicitly reported in App.~\ref{app:Simulation}).
In the following results, we assumed $\gamma_{TR,d}/2\pi =100$ kHz as realistic transmon dephasing rate, and checked the performances for different values of the nanomechanical qubits $T_2$ times.  

As a first example, we show in Fig.~\ref{fig:simulator}(a) the quantum simulation of a spin-1 Hamiltonian initialized in a fully polarized eigenstate and experiencing tunneling of the magnetization. 
The simulation of Hamiltonians involving $S>1/2$ spins can be performed by encoding the state of each spin-$S$ into that of $2S$ qubits.
The target $S=1$ Hamiltonian in this case reads
\begin{equation}
\mathcal{H}_{S1} = D S_z^2 + E (S_x^2 - S_y^2 ) \, .
\label{eq:HS1}
\end{equation}
By considering the total spin as a combination of two $1/2$ spins, $S_{\alpha}=s_{\alpha,1}+s_{\alpha,2}$, the mapped Hamiltonian 
$\tilde{\mathcal{H}}_{S1} = 2D s_{z,1}s_{z,2} + 2E (s_{x,1}s_{x,2}  - s_{y,1} s_{y,2}) \, $ results in a sum of two body-terms that can be easily implemented in our platform. In particular, the evolution induced by $s_{x,1}s_{x,2}  - s_{y,1} s_{y,2}$ is obtained by two $R_y(\pi)$ rotations on one of the two qubits, preceding and following the two-qubit evolution provided by $H_{eff}$ (see App.~\ref{app:Simulation}).
The exact time evolution of the total magnetization, $\langle S_z \rangle$, is compared in fig. \ref{fig:simulator}(a) to a digital quantum simulation of $\tilde{\mathcal{H}}_{S1}$ with our electromechanical set-up, either for $\gamma_{NR,d} =0$ or $\gamma_{NR,d}/ 2\pi =1$ kHz. The overall quantum simulation works very well with an average fidelity $\mathcal{F}=0.999$ for $\gamma_{NR,d} =0$ and $\mathcal{F}=0.988$ for $\gamma_{NR,d}/ 2\pi =1$ kHz, respectively. 

As a further test, Fig.~\ref{fig:simulator}(b) reports the digital quantum simulation of the total magnetization along $x$, i.e. $\langle S_x \rangle=\mathrm{Tr}[\rho(s_{x,1} + s_{x,2})]$ \cite{Sx}, for the transverse field Ising model of two 1/2 spins {(recently become the subject of intense theoretical activity in the context of analog quantum simulations\cite{Tian2010,Viehmann2013,You2013,Du2015})}, which reads
\begin{equation}
\mathcal{H}_{TIM} = \Lambda s_{x,1} s_{x,2}  + b (s_{z,1} + s_{z,2})  \, ,
\label{eq:H_TIM}
\end{equation}
where we set $\Lambda = 2 b = \Gamma$  for the specific simulation in Figure. 
Notice that the computation of each point in Fig.~\ref{fig:simulator}(b) (with $N=10$ Trotter steps) requires the sequential concatenation of 20 two-qubit gates and 40 single-qubit rotations (each one operated in parallel on both qubits). Although this makes the simulation much more demanding, 
the fidelity is already $0.90$ for $\gamma_{NR,d}/ 2\pi =1$ kHz, which steeply increases to $0.96$ if a more optimistic  $\gamma_{NR,d}/ 2\pi =100$ Hz is assumed. The latter result is especially noteworthy if one considers the total computational time required for the longest sequence of gates (i.e.,  corresponding to the last point in the Figure), which is about $150$ $\mu$s. This confirms the robustness and potential strength of the quantum computing platform introduced in this work. \\
{ In addition, by exploiting generalized Jordan-Wigner transformations to map fermionic into spin operators \cite{Laflamme,Bari,Santini2011}, it is possible to simulate many-body fermionic systems. 
In particular, our effective qubit-qubit interaction already implements a XY model, which is the essential building-block in the simulation of hopping processes in fermionic Hamiltonians \cite{Chiesa2015,Reiner2016}. }

\section{Conclusions}
In conclusion, we have proposed a scalable architecture to realize an electromechanical digital quantum simulator, based on state-of-the-art technology.
Qubits are encoded in the anharmonic vibrational modes of mechanical nanoresonators, whose coupling is mediated by virtual excitations of an auxiliary transmon and can be switched on and off by tuning their resonance frequency. We have shown that the fidelity of elementary gates is practically unaffected by the transmon decoherence and remains remarkably high even with the inclusion of realistic values of the nanoresonators decoherence rates. These elementary gates are concatenated into quantum simulation algorithms and very good results are found for the implementation of non-trivial models, such as the transverse field Ising model and the XY model.

\acknowledgements
This work was partly funded by the Italian Ministry of Education and Research (MIUR) through PRIN Project 2015 HYFSRT  ``Quantum Coherence in Nanostructures of Molecular Spin Qubits''. 
A. C. acknowledges financial support from ``Fondazione Angelo Della Riccia". The authors acknowledge useful discussions with F. Troiani. \\
$^+$ These authors contributed equally to this work.

\appendix

\section{Anharmonicity of nanomechanical resonators} 
\label{app:NonLin}

\begin{figure*}[t]
\centering
\includegraphics[scale=0.5]{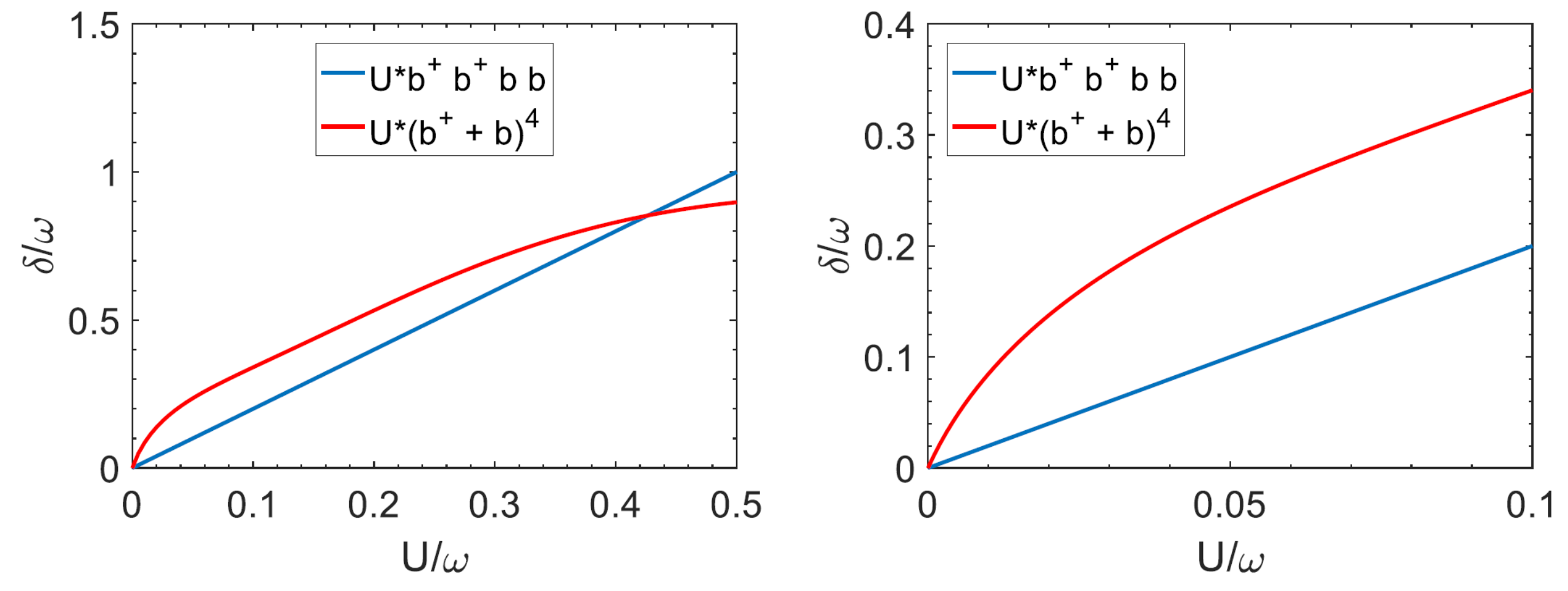} 
\caption{Comparison between two possible models for the single-phonon nonlinearity of the nanomechanical resonators. Panel on the right focuses on the region of interest for our setup.}
\label{fig:NlModels}
\end{figure*}

A certain degree of anharmonicity is the essential ingredient that allows a faithful encoding of information on the ground and first excited levels of each nano-electromechanical resonator. In our theoretical description we assumed a very simple model for the non linear contribution to the energy spectrum: indeed, a diagonal  shift of the $|1\rangle\leftrightarrow |2\rangle$  transition with respect to $|0\rangle\leftrightarrow |1\rangle$ (in the Fock number basis representation of each qubit) already contains all the relevant features, while keeping the description easy to understand and analytically transparent. Here we compare this simplified description with a more realistic and commonly used model of anharmonic mechanical oscillators, with the aim of better clarifying the physical properties and parameters that are required from our nano-electromechanical devices.

From the perspective of the total bosonic Hilbert space (\textit{i.e.}\ without truncation on the maximum number $n_{1,\mathrm{max}}$ and $n_{2,\mathrm{max}}$ of allowed excitations in each resonator), the diagonal nonlinearity model can be written as a Kerr-type Hamiltonian, i.e.
\begin{equation}
H_{nl,diag} = U b^\dagger b^\dagger b b \, .
\label{eq:HnlDiag}
\end{equation}
On the other hand, a widely accepted model for nonlinear nanomechanical resonators is rather given by 
\begin{equation}
H_{nl} = U (b^\dagger+b)^4 \propto \hat{x}^4 \, .
\label{eq:Hnl}
\end{equation}
As a first step in comparing the two models, we will now show that the first one, which we have adopted in this work, underestimates the degree of required nonlinearity given the same parameter $U$ in the range of interest, \textit{i.e.}\ it predicts a smaller shift $\delta = \omega_{21}-\omega_{10}$, with respect to \eqref{eq:Hnl}. This is easily seen in Fig.~\ref{fig:NlModels}, where we compute the shift $\delta$ by performing a numerical diagonalization of both models, after adding the free Hamiltonian $H_0 = \omega b^\dagger b$ and setting $n_{max} = 10$. In the plot, $U$ and $\delta$ are both expressed as fractions of the bare frequency of the oscillator $\omega$. As it is evident from the results, in the range that corresponds to the region of interest for our purposes ($U/\omega \leq 0.1$), the model Hamiltonian \eqref{eq:HnlDiag} that we employed is always quite conservative in terms of quantitative estimation of the nonlinear single-phonon contribution.

In addition to the eigenvalues, we also compared the eigenvectors corresponding to the first three energy levels (namely the computational basis plus the first extra level) for the two models. In this case, we selected the parameter $U$ in two different ways such that the gap, $\delta$, is the same both for $H_{nl,diag}$ and $H_{nl}$. 
The two bare frequencies were assumed as $\omega_1  = 85\,$MHz and $\omega_2  = 75\,$MHz, corresponding to the values used in the simulations shown in the main text. The following table summarizes the fidelity, $\mathcal{F}$, of the eigenvalues $|n\rangle_{nl}$ obtained from \eqref{eq:Hnl} as compared to the corresponding bare Fock state, $|n\rangle$.

\begin{table}[!h]
\centering
\begin{tabular}{|c|c|c|}
\hline
$n$ & $\mathcal{F}(\omega = \omega_1)$ & $\mathcal{F}(\omega = \omega_2)$ \\ 
\hline
$1$ & $0.9996$ & $0.9995$ \\ 
$2$ & $0.9971$ & $0.9963$ \\ 
$3$ & $0.9899$ & $0.9872$ \\ 
\hline
\end{tabular}
\end{table}

\noindent By using the $|n\rangle_{nl}$ states as elements of the computational basis, the very same protocol machinery that we presented in the paper can be used to implement single-qubit rotations and the $\sqrt{i\text{\textsc{SWAP}}}$ gate. Indeed, electrical pulses can still be used to tune the fundamental transition frequency, $\omega_{01}$, for both oscillators, thus bringing them to resonance when needed. Moreover, the operators $b_i$ and $b_i^\dagger$ promote transitions between the new eigenvectors, albeit with a slightly different matrix element $X_{kl} = \langle k | b |l\rangle$, as summarized in the following table.

\begin{table}[!h]
\centering
\begin{tabular}{|c|c|c|c|}
\hline
$X_{kl}$ & Fock states & $|n\rangle_{nl}$ for $(\omega = \omega_1)$ & $|n\rangle_{nl}$ for $(\omega = \omega_2)$ \\ 
\hline
$X_{01}$ & $1$ & $1.0005$ & $1.0007$ \\  
$X_{12}$ & $\sqrt{2}$ & $1.4170$ & $1.4179$ \\ 
\hline
\end{tabular}
\end{table}

\noindent The Hamiltonian \eqref{eq:Hnl} only couples Fock states differing by an even number of excitations. This means that $|n\rangle_{nl}$ and $|n+1\rangle_{nl}$ are still orthogonal to each other, since they are superpositions of even or odd Fock states only, namely eigenvectors of the parity operator on the Fock basis belonging to different eigenspaces. As a consequence, the matrix element $X_{nn}$ vanishes. The same is not true for $X_{n(n+2)}$, meaning that $b_i$ and $b_i^\dagger$ could in principle promote $|0\rangle_{nl}\leftrightarrow |2\rangle_{nl}$ transitions outside the computational basis, \textit{e.g.}\ during the single qubit $xy$ rotations. However, in our protocol this effect remains negligible in view of the large energy gap between the two eigenstates. 

\begin{figure}[b]
\centering
\includegraphics[scale=0.4]{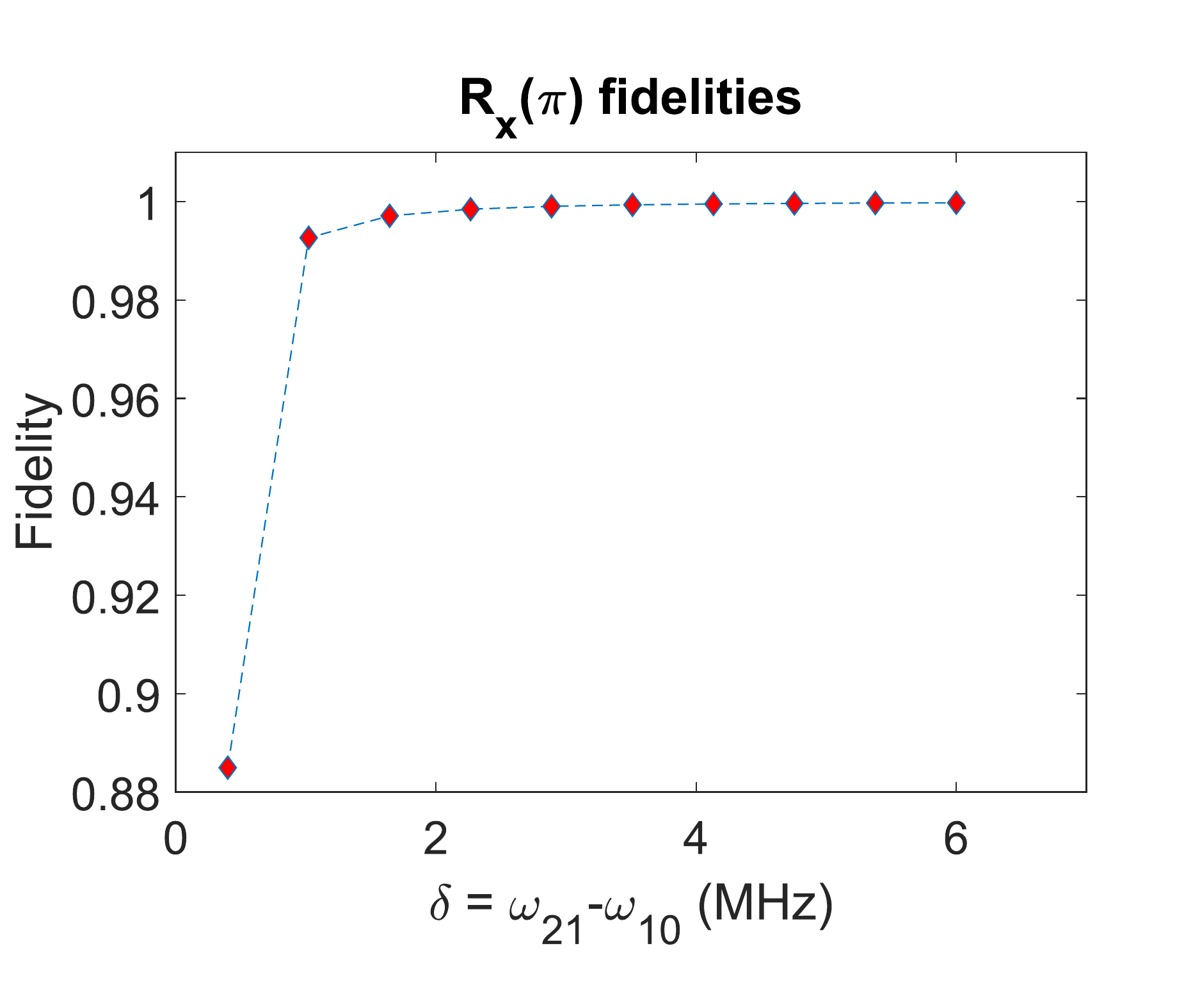}
\caption{Fidelity of single quibit $R_x(\pi)$ gate as a function of the nonlinear shift between the first and second energy gaps. Gaussian tuning pulses were used (see App.~\ref{app:Gates}) to activate and control the gates.}
\label{fig:RxVsU}
\end{figure}

In conclusion, the only significant consequence that we must keep into account is the rescaling of the coupling elements $g_i \rightarrow g_i X_{xy}$ (notice that, in principle, this is specific for every transition) and the shortening, by a factor $X_{01}$, of the time required for the single qubit $xy$ rotations. Numerical simulations of individual gates within this framework show behaviors and fidelities that are very close (the difference is below $0.1\%$) to the simplified case that we have adopted for the numerical simulations. \\

\textit{Fidelity vs nonlinear shift}. We evaluated the relevance of the nonlinearity to the quantum information processing tasks by performing the fundamental gates for different values of the parameter $\beta$ used in the main text. As an example, in Fig.~\ref{fig:RxVsU}, we show the data points obtained for a single qubit $R_x(\pi)$ rotation as a function of the nonlinear gap difference $\delta = \omega_{21}-\omega_{10}$. Single qubit rotations are the most sensitive to non sufficient of nonlinearity in our scheme, since the total number of excitations is not conserved by the external pulse proportional to $b+b^\dagger$, in general. The simulations were carried out with the same nonlinearity model and parameters as in the main text but varying $\beta$, in the absence of external dissipations. The most significant feature is the large plateau very close to unity for almost all values above $\delta \simeq 2\,$MHz. Moreover, we find that a nonlinear shift $\delta = 1\,$MHz is already sufficient to give theoretical fidelities $\mathcal{F}>0.99$, which is the threshold required on single gates to successfully apply error correction protocols such as the surface code \cite{Barends2014}.  \\

\begin{figure*}
\centering
\includegraphics[scale=0.46]{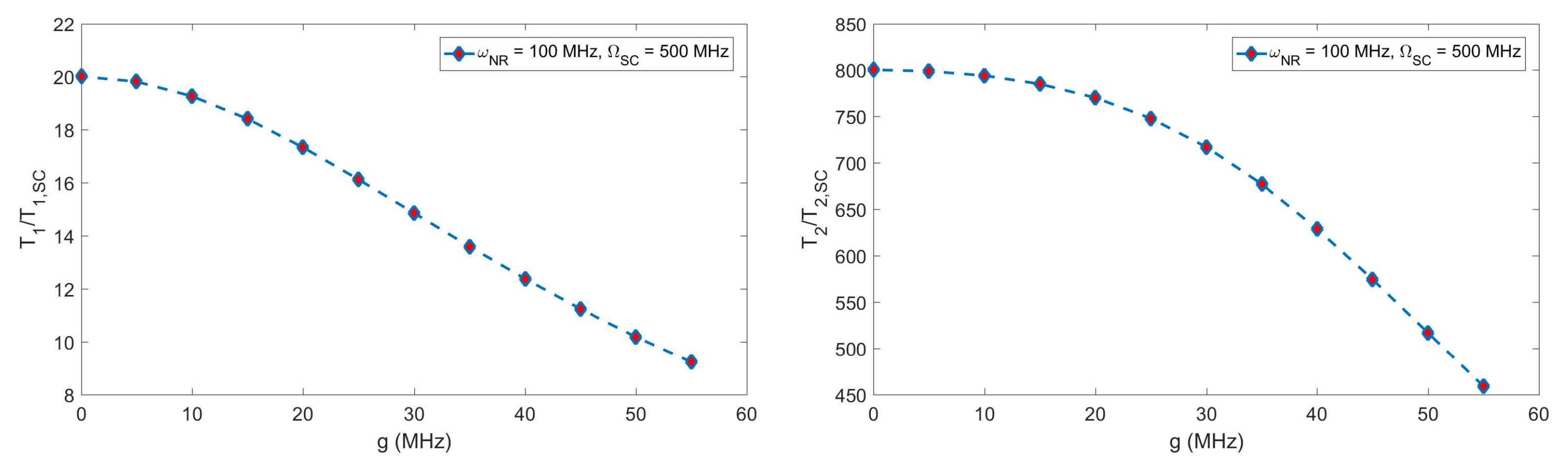} 
\caption{Dissipation and coherence times for the coupled nanomechanical resonator and superconducting circuit system, obtained from Bloch-Redfield simulations. The values of overall $T_1$ and $T_2$ timescales are expressed in units of the corresponding {typical} timescales for the superconducting circuit assumed in the simulation, as taken from the literature (i.e., ${T}_{1,SC} = 1\,$ms and ${T}_{2,SC} \simeq 10\,\mu$s).}
\label{fig:T1T2}
\end{figure*}

\section{Decoherence effects of phonon nonlinearity induced by a superconducting circuit} 
\label{app:Single-Ph-Nonlin}

With reference to the scheme introduced in Sec.~\ref{sec:mech_anharm}, and to show its effectiveness in defining efficient nanomechanical qubits, we have quantitatively analyzed the actual effect of introducing  the SC element in terms of dissipation ($T_1$) and coherence times ($T_2$) of the collective excitations, by applying the formalism of open quantum systems. In view of the strong internal coupling ($g$ can be a significant fraction of $\omega_{NR}$) and of the presence of non-negligible counter-rotating terms, a Bloch-Redfield master equation with suitable secular approximation \cite{Breuer2002} was used instead of a more phenomenological collection of Lindblad terms derived for the bare nanomechanical resonator and SC subsystems. In a nutshell, the Bloch-Redfield approach starts by diagonalizing the full system Hamiltonian, a procedure that is indeed consistent with the fact that our qubits are actually defined as slightly mixed excitations, and then derives the dissipation and pure dephasing terms from the matrix elements, computed between the true system eigenstates, of a set of operators describing the coupling to the environment. In our case, we worked with the following model
\begin{equation}
H_{TOT} = H_{NR+SC} + H_{env,NR} + H_{env,SC} + H_{I} \, ,
\end{equation}
where $H_{NR+SC}$ is given in Eq.\ \eqref{eq:H_NRSC}, $H_{env,i} = \sum_k \omega_{k,i} d^\dagger_{k,i} d_{k,i}$ is a collection of harmonic bath modes ($i = NR, SC$) and the system-bath coupling $H_I$ is a sum of terms of the form
\begin{equation}
H_{I,S} = O \otimes \sum_k g'_{k,i} (d_{k,i} + d_{k,i}^\dagger) \, .
\end{equation}
Here $O$ is a hermitian system operator describing individual interaction mechanisms for either the nanoresonator or the superconducting circuit. We use, for example, the operators $O_{NR} = \{b+b^\dagger;b^\dagger b\}$ and $O_{SC} = \{ \sigma_- +\sigma_+;\sigma_z\}$ corresponding to dissipative and dephasing processes on the bare nanoresonator and SC respectively. The resulting master equation will correctly describe all possible transitions induced by the environment on the effective dynamics of the coupled nanoresonator and SC system, with rates that are proportional to the spectral functions of the environment evaluated at the relevant transition frequencies. We specify such functions as zero temperature white noise spectra
\begin{equation}
S_i (\omega) = \begin{cases} \gamma_{i,d} \quad & \text{for } \omega = 0 \\
\gamma_{i} \quad & \text{for }\omega > 0 \\
0 \quad & \text{for }\omega < 0 \\
 \end{cases}
\end{equation}
where, again, $i = NR,SC$ and $\gamma$ ($\gamma_d$) represent dissipation (pure dephasing) contributions in the uncoupled case. These numerical simulations were carried out with the QuTiP library in Python\cite{Qtip} (see also http://qutip.org/). In Fig.~\ref{fig:T1T2} we show the results for the change in the total $T_1$ (decay of diagonal terms) and $T_2$ (decay of coherences) of the coupled system as a function of $g$, obtained by observing the time evolution of an initial superposition of the computational basis elements. In practice, the data are obtained by fitting the exponential decay of the excited state occupation probability and of the off-diagonal element (coherence) of the resulting density matrix. The following parameters were used in these simulations: $\omega_{NR} = 100$ MHz, $\Omega_{SC} = 500$ MHz, $\gamma_{NR}  = 50$ Hz (corresponding to $T_{1,NR} = 20$ ms), $\gamma_{NR,d} = 200$ Hz (which results in $T_{2,NR} = 8$ ms), $\gamma_{SC} = 1$ kHz, and $\gamma_{SC,d} = 50$ kHz. 
In particular, these parameters correspond to uncoupled values of {$\text{T}_{1,SC} = 1\,$ms and $\text{T}_{2,SC} \simeq 10\,\mu$s}  for the low-frequency SC element, as experimentally reported, e.g., in Ref. \onlinecite{Pop2014}. 
As it can be appreciated from the plots, the additional superconducting element is predicted to affect the original performances of the nanomechanical oscillator by less than an order of magnitude, thus still preserving a significant advantage over the typical dissipation and coherence times of, e.g., transmon qubits or Cooper pair box with a large charging energy used as the SC element in this calculations, whose $T_1$ and $T_2$ values are taken as a reference for normalization in the figures. 
Before concluding, we also notice that the introduction of a superconducting nonlinear element is in principle compatible with the parallel use of other strategies designed to enhance the single-phonon nonlinearity, such as the external static electric fields or clamping techniques mentioned in the main text.

\begin{figure*}[t]
\centering
\includegraphics[width=0.70\textwidth]{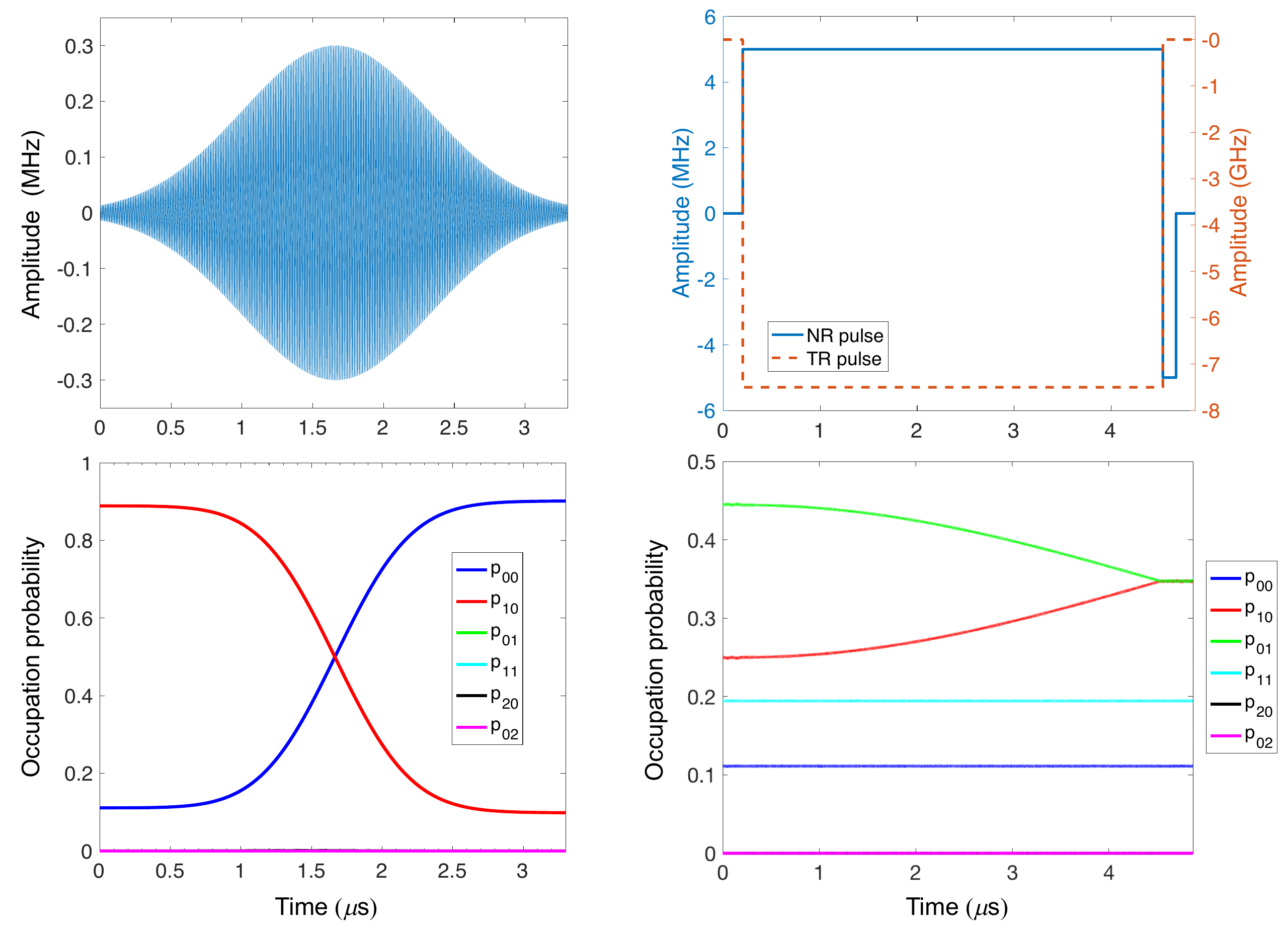} 
\caption{Numerical simulations of single- and two-qubit quantum gates in an electromechanical circuit. Here, the system undergoes unitary evolution, with $\omega_1 = 85\,$MHz and $\omega_2 = 75\,$MHz. With $p_{ab}$ we denote the component of the two-qubit wavefunction on the corresponding Fock state $p_{ab}=\left|\langle ab|\psi\rangle\right|^2$. (Bottom left) $R_x(\pi)$ rotation performed on qubit 1 while qubit 2 is kept isolated and the transmon frequency $\Omega = 10\,$GHz is left unchanged. (Top Left) The gaussian oscillating pulse acting on qubit 1 needed for the $R_x(\pi)$ gate. The peak amplitude is $0.3\,$MHz. (Bottom right) A typical time evolution displaying a $\sqrt{\text{iSWAP}}$ operation, with a short idle phase before and after the gate; (Top right) Frequency shifts operated on qubit 2 and on the transmon during the time evolution, including the rephasing stage on the qubit. Notice that the renormalization shift $\lambda_2$ on the qubit is not included.}
\label{fig:Gates}
\end{figure*}

\section{Single- and two-qubit gates}
\label{app:Gates}

As we have pointed out in the main text, the fundamental ingredient needed to perform single and two qubit gates is the possibility to dynamically tune the oscillation frequency of the nanomechanical resonators. Here we show in full detail the protocols that are needed to obtain two specific gates, namely a single qubit $x$-rotation of an angle $\alpha$ and a two-qubit $\sqrt{\text{i\textsc{SWAP}}}$ gate, ideally defined by the unitary matrix

\begin{equation} \label{iswap_gate}
U_{\sqrt{\mathrm{iSWAP}}} = 
\left(\begin{array}{cccc} 1 & 0  &  0  &  0 \\ 
                                      0 & 1/\sqrt{2} & i/\sqrt{2}   & 0 \\  
                                      0 & i/\sqrt{2} & 1/\sqrt{2}   & 0 \\
                                      0 & 0 & 0   & 1 \\
                                      \end{array}\right) \,\,\,\, .
\end{equation}

Unless explicitly stated, we will assume throughout the discussion that the permanent renormalization shifts (encoded in the parameters $\lambda_i$ of the main text) have already been taken into account.\\
{We adopt a modeling similar to Ref. \onlinecite{Rips2013}, where diagonal terms of the form
\begin{equation}
H^z_i(t) = V^z (t) x_i^2 = V^z (t) (b_i + b_i^\dagger)^2
\end{equation} 
modify the oscillation frequency, while transverse components
\begin{equation}
H^{xy}_i(t) = V^{xy} (t) x_i = V^{xy} (t) (b_i + b_i^\dagger)
\end{equation}
displace the equilibrium position and can be used for $x$- and $y$-rotations. \\
$R_z$-rotations can be performed by changing the qubits oscillation frequency by an amount $\delta\omega$ for a time interval $\delta t$. 
To implement it, a step-like pulse can be used, the temporal switching of the external fields being only limited by the response time of the control electronics (typically in the ns timescale). The resulting $H_i^z(t) = \delta\omega \Theta_{t_0}(\delta t) b_i^\dagger b_i$ will produce a phase $\phi_z = -\int dt \delta\omega \Theta_{t_0}(\delta t) = - \delta\omega \delta t$ on the $|1\rangle$ component of the basis. With $\Theta_{t_0} (\delta t)$ we denoted a unitary step function starting at $t_0$ with duration $\delta t$. }\\
As a specific example, a $R_x(\alpha)$ gate on the first nanoresonator is obtained by applying a transverse pulse
\begin{equation}
H^{xy}_1(t) = V^{xy} (t) x_1 = V^{xy} (t) (b_1 + b_1^\dagger)
\end{equation}
with
\begin{equation}
V^{xy} (t) = A(t,t_0,\sigma)V^{xy}_0 \cos (\omega_1 t) \, .
\end{equation}
Here $V^{xy}_0$ denotes the amplitude scale of the pulse, while $A(t,t_0,\sigma)$ is a time-dependent modulation of the oscillatory part that describes the on/off dynamics of the gate. For example, a square pulse $A(t,t_0,\sigma) = \Theta(\sigma/2-|t-t_0-\sigma/2|)$ starting at $t_0$ and lasting for $\sigma = \alpha/V^{xy}_0$ ($\Theta$~is the unit step function) is one of the possible choices for the envelope function. Given that we are not dealing with real two-level systems, but rather trying to restrict the dynamics of a nonlinear harmonic oscillator to the first two levels, the choice of the pulse profile is of great importance and can lead to significant improvement of the performances. Indeed, the frequency spectrum of a cosine-like function modulated by a square pulse may not be sufficiently narrow around the target $\omega_1$ to avoid the activation of unwanted transitions that are close in energy, \textit{e.g.}\ the $|1\rangle \leftrightarrow |2\rangle$ transition. An easy to implement but powerful tool in this context is provided by gaussian pulses, which give a fast decaying gaussian frequency spectrum. The envelope can be chosen as
\begin{equation}
A(t,t_0, \sigma) = e^{-\frac{(t-t_0)^2}{2\sigma^2}}
\end{equation}
with $\sigma = \alpha/(\sqrt{2\pi} V^{xy}_0)$. The gate lasts approximately $2.5$-$3\sigma$ on both sides of the central peak at $t=t_0$. By using gaussian pulses, the amount of nonlinearity $U$ that is required to obtain reasonably large fidelities can be reduced with respect to the square pulse case, or alternatively the strength of the tuning $V^{xy}_0$ can be increased, thus diminishing the total gating time.

For the two-qubit gate the transmon-mediated interaction must be activated. Good isolation of the qubits during the idle phase and single-qubit gates requires that they are detuned from each other and that the transmon is at a sufficiently high frequency (\textit{e.g.}\ $\Omega\simeq10\,$GHz) to strongly suppress the residual effective coherent coupling, whose strength is
\begin{equation}
\Gamma = \frac{4 g_1 g_2 \Omega(\omega_1^2 + \omega_2^2 -2\Omega^2)}{(\Omega^2 - \omega_1^2)(\Omega^2 - \omega_2^2)}
\label{eq:Gamma}
\end{equation}
In principle, tuning the two qubits to resonance (\textit{e.g.}\ to a common intermediate frequency $\omega_r = |\omega_1 - \omega_2|/2$) is sufficient to activate coherent oscillations: the desired $\sqrt{\text{i\textsc{SWAP}}}$ gate is then obtained for an interaction time $\tau = \pi/|\Gamma|$. In this case, square pulses already give good fidelities in the range of parameters that we studied. The interaction is then turned off by shifting back the frequencies of the nanoresonators, and two rephasing single qubit $z$-rotations are applied to correct for the additional phase accumulated by the qubits with respect to their evolution with the original bare $\omega_{1,2}$. This latter step is obtained by an inverse $-\xi_i = \omega_i - \omega_r$ pulse lasting for a time $\tau'= \operatorname{mod}(\tau ,2\pi)$. In order to shorten the required gating time, one can exploit one peculiar property of our setup, namely the possibility of dynamically tuning the transition frequency $\Omega$ of the transmon. This is, of course, of great importance when performing long sequences in the presence of realistic dissipation processes. This shift, which is implemented in practice by varying the magnetic flux concatenated with the transmon, is essentially just another time-dependent contribution to the Hamiltonian in the interaction picture
\begin{equation}
H_{TR}(t) = \delta\Omega(t) \frac{\sigma_z}{2}
\end{equation}
Changing the frequency of the transmon affects all the effective qubit-qubit Hamiltonian parameters: in particular, reducing $\Omega$ (and thus the detuning $\Delta$ with respect to the nanoresonators) increases the coupling $\Gamma \propto g^2/\Delta$ and modifies the renormalizations $\lambda_i$. Needless to say, this procedure is limited both by the tunability range of the transmon and by the validity of the perturbative expansion in terms of $g^2/\Delta$. Given some values of the external dissipation rates, for example, there exists an optimal $\delta\Omega$ that increases the gate fidelity without loosing too much of the agreement between the actual behavior of the system and what is expected from the effective Hamiltonian description. In our protocols, we set a non zero $\delta\Omega$ during the coherent interaction time $\tau$ (which must then be computed by using $\Omega+\delta\Omega$ in equation \ref{eq:Gamma}) and we put the transmon back to its original frequency already during the rephasing stage. It is worth noting explicitly that when the transmon frequency is modified, the permanent shifts $-\lambda_i$ applied to the qubits must be adjusted accordingly. Fig.~\ref{fig:Gates} shows a single qubit $R_x(\pi)$ rotation and a two-qubit gate together with the required pulse sequences.

\section{Derivation of the effective qubit-qubit interaction}
\label{app:Heff}

We provide here a detailed derivation of the effective Hamiltonian describing the interaction between the two qubits, mediated by virtual fluctuations of the interposed transmon. 
This is obtained by considering 
\begin{equation}
H_{int} = \sum_{i=1}^2 g_{i,x} \left(b_i+b_i^\dagger\right)\sigma_x 
\end{equation}
as a weak perturbation with respect to
\begin{equation}
H_{0} = \sum_{i=1,2} \left[ \omega_i b^\dagger_i b_i + \beta b^\dagger_i b^\dagger_i b_i b_i \right] +  \frac{\Omega} {2}\sigma_z .
\end{equation}
This condition is ensured, provided that $g_i \ll \Delta_i = \omega_i-\Omega$. Then we eliminate the transmon degrees of freedom by second order expansion and restrict to the two-qubit computational basis, where the transmon is frozen in its ground state. 
In this subspace, the matrix elements of the effective Hamiltonian are given by:
\begin{widetext}
\begin{equation}
\begin{split}
\left\langle \mu \nu | \tilde{H}_{eff} | \mu^\prime  \nu^\prime \right\rangle & = \frac{1}{2}\sum_{m_1,m_2} \left\langle n_1 \downarrow n_2 | H_{int} | m_1 \uparrow m_2 \right\rangle \left\langle m_1 \uparrow m_2 | H_{int} | n_1^\prime \downarrow n_2^\prime \right\rangle  \\
& \times \left[ \frac{1}{\omega_1 (n_1-m_1) + \omega_2 (n_2-m_2)- \Omega} + \frac{1}{\omega_1 (n_1^\prime-m_1) + \omega_2 (n_2^\prime-m_2)-\Omega }  \right].
\end{split}
\end{equation}
\end{widetext}
Here $| n_1 \sigma n_2 \rangle $ are states in the full Hilbert space, with $n_i$ being the bosonic occupation of each nonoresonator mode and $\sigma = \downarrow$ $(\uparrow)$ indicating the ground (excited) state of the transmon. 
Conversely, $\tilde{H}_{eff}$ operates in the 2-qubits Hilbert subspace, spanned by the computational basis $|\mu \nu\rangle$, with $\mu,\nu = 0,1$. 
The sum runs over all the states with $\sigma = \uparrow$ (excited transmon). Finally, $\tilde{H}_{eff}$ is decomposed in terms of Pauli operators $\sigma_\alpha^i$. 
We report here the form it takes if bosonic states up to $m_i=2$ are included:
\begin{equation}
\tilde{H}_{eff} = \sum_{i=1}^2 \left(\frac{\lambda_i}{2}\sigma^i_{z}\right) + \frac{\Gamma}{4} \sigma^1_x\sigma^2_x + \text{const.}, 
\label{Heff}
\end{equation}
where
\begin{equation}
\Gamma = \frac{4 g_1 g_2 \Omega(\omega_1^2 + \omega_2^2 -2\Omega^2)}{(\Omega^2 - \omega_1^2)(\Omega^2 - \omega_2^2)} \, 
\end{equation}
is the effective coupling, while
\begin{equation}
\lambda_i =  \frac{-2g_i^2(\Omega^2+\omega_i(2\beta + \Omega))}{     (2\beta+\Omega+\omega_i)(\Omega^2 - \omega_i^2)}
\end{equation}
are single-qubit energy shifts (corresponding to a renormalization of the qubit frequencies). 
Notice that $\tilde{H}_{eff}$ was obtained while truncating bosonic occupancy up to 2, and it was restricted to the computational basis only afterwards.\\
In order to get rid of the frequency renormalization induced by the transmon, 
we shift the bare frequencies $\omega_i$ by an amount $-\lambda_i$ throughout the whole gating dynamics. From the theoretical point of view, this detuning is implemented by introducing  additional terms $-\lambda_i b_i^\dagger b_i$ in the model Hamiltonian.\\
It is worth noting that the effective interaction in Eq.~\eqref{Heff} can be decomposed in two terms:
\begin{equation}
\frac{\Gamma}{4} \sigma^1_x\sigma^2_x = \frac{\Gamma}{8} \left(\sigma^1_x\sigma^2_x + \sigma^1_y\sigma^2_y\right) + \frac{\Gamma}{8} \left(\sigma^1_x\sigma^2_x - \sigma^1_y\sigma^2_y\right)
\end{equation}
The first term on the r.h.s. is an effective XY interaction that couples the $|10\rangle$ and $|01\rangle$ elements of the computational basis. The second term is generated by the counter-rotating contributions in the original interaction Hamiltonian, and couples the $|00\rangle$ and $|11\rangle$ components: given the large energy gap $ \simeq \omega_1 + \omega_2 \gg \Gamma$ between them, this part is actually negligible. Hence, we can re-write the effective Hamiltonian as reported in the main text:
\begin{equation}
H_{eff} = \sum_{i=1}^2 \left(\frac{\lambda_i}{2}\sigma^i_{z}\right) + \frac{\Gamma}{8} \left( \sigma^1_x\sigma^2_x +\sigma^1_y\sigma^2_y \right) + \text{const.}
\label{eq:Heff}
\end{equation}
The second term of $H_{eff}$ is practically ineffective as far as $|\omega_1-\omega_2| \gg \Gamma$ and it is only switched on when the two qubits are brought in resonance. In that case, the matrix form of the time-evolution operator
$U_{XY} (t) = e^{-i \frac{\Gamma}{8} \left( \sigma^1_x\sigma^2_x +\sigma^1_y\sigma^2_y \right)  t}  $ expressed in the computational basis $\{|00\rangle, |10\rangle,|01\rangle, |11\rangle\}$ is
\begin{equation}
U_{XY} (t) = 
\left(\begin{array}{cccc} 1 & 0  &  0  &  0 \\ 
0 & \text{cos}\frac{\Gamma t}{4} & -i~\text{sin}\frac{\Gamma t}{4}   & 0 \\  
0 & -i~\text{sin}\frac{\Gamma t}{4} & \text{cos}\frac{\Gamma t}{4}   & 0 \\
0 & 0 & 0   & 1 \\
\end{array}\right),
\end{equation}
which corresponds to the $\sqrt{\text{iSWAP}}$ gate, ideally Eq.~\eqref{iswap_gate}, if we choose to stop the evolution at $t = \pi/|\Gamma|$.

\section{Residual thermal occupancy of nanomechanical resonators}
\label{app:Thermal}

In our analysis, we assumed that the system can be cooled at sufficiently low temperatures to safely neglect the effect of thermal interaction with the environment. Examples of ground state cooling of nanomechanical systems are already known in the literature. However, this is still a challenging issue from the experimental point of view, especially when resonators in the MHz range are considered. For this reason, we studied how the fidelities of the gates change if we add, on top of all the dissipation mechanisms that we already considered, a residual thermal interaction between the nanoresonators and the surrounding environment. Since the transmon has a much higher transition frequency (\textit{i.e.}\ an higher temperature is sufficient to cool it in the ground state), we did not include thermal noise acting on this element. We used the standard Lindblad terms to model the interaction of the $i$-th oscillator with a bosonic thermal reservoir
\begin{widetext}
\begin{equation}
\mathcal{L}_i [\rho] = \frac{\chi}{2}\left[\bar{n}\left(\omega_i,T\right)+1\right]\left(2b_i\rho b_i^\dagger - b_i^\dagger b_i\rho - \rho b_i^\dagger b_i\right) + \frac{\chi}{2}\bar{n}(\omega_i,T)\left(2b_i^\dagger\rho b_i - b_i b_i^\dagger\rho - \rho b_i b_i^\dagger\right) 
\end{equation}
\end{widetext}
where ($\hbar = k_B = 1$)
\begin{equation}
\bar{n}(\omega_i, T) = \frac{1}{\exp\left(\frac{\omega_i}{T}\right)-1}
\end{equation}
In our simulations, we chose $\bar{n} = 0.1$ for both oscillators, as this generally represents a good estimate of the achievable residual thermal occupation. The rate $\chi$ can be inferred from typical line broadening of the nanoresonators, which is of the order of few tens of Hz. When average values for the other dissipation mechanism are taken into account, the effect of the residual thermal occupation is of the order of $0.01-0.1\%$ of the single gate fidelity if $\chi = 50\,$Hz, and of $1\%$ if we use the very pessimistic value $\chi = 1\,$kHz. The overall effect is therefore comparable to the general (zero temperature) dissipation mechanism that we included via the rates $\gamma_i$.

\section{Quantum simulation of generic two-spin interactions}
\label{app:Simulation}

The time evolution \begin{equation}
U_{XY}(t)= \text{exp} \left[-i J \left( \sigma^1_x\sigma^2_x +\sigma^1_y\sigma^2_y \right)  t \right]
\end{equation}
induced by $H_{eff}$ [Eq. (\ref{eq:Heff})] can be mapped into a generic spin-spin evolution
\begin{equation}
U_{\alpha\beta} (t) = e^{-i J \left( \sigma_{\alpha}^1 \sigma_{\alpha}^2  +  \sigma_{\beta}^1 \sigma_{\beta}^2\right)t},
\end{equation}
by properly combining it with single-qubit rotations. 
For instance, the following identities hold (see also Ref. \onlinecite{LasHeras2014}):
\begin{equation}
U_{XZ}(t) = R_x^{12}\left(\frac{\pi}{2}\right) U_{XY}(t) R_x^{12}\left(\frac{-\pi}{2}\right)  
\end{equation}
and
\begin{equation}
U_{YZ}(t) = R_y^{12}\left(\frac{\pi}{2}\right) U_{XY}(t) R_y^{12}\left(\frac{-\pi}{2}\right)  
\end{equation}
Here $R_\alpha^{12} (\theta) = \text{exp}\left[-i\left(  \frac{\sigma_{\alpha}^1}{2}+\frac{\sigma_{\alpha}^2}{2} \right) \theta \right] $ is a simultaneous rotation of both qubits by an angle $\theta$ about $\alpha$ axis. \\
Another useful identity is
\begin{equation}
U_{XY}(t)^- = R_y^{1}\left(\pi\right) U_{XY}(t) R_y^{1}\left(-\pi\right)  ,
\end{equation}
with $U_{XY}^-(t) = \text{exp} \left[-i J \left( \sigma_{x}^1 \sigma_{x}^2  -  \sigma_{y}^1 \sigma_{y}^2\right)t \right] $ and $R_\alpha^{j} (\theta) = \text{exp}\left[-i  \frac{\sigma_{\alpha}^j}{2} \theta \right] $.
Note that the combination of $U_{XY}(t)$ with $U_{XY}^-(t)$ was used to simulate the Ising XX interaction reported in the main text:
\begin{equation}
U_{XX}(t) = U_{XY}(t) U_{XY}^-(t) .
\label{eq:XX}
\end{equation}
Here $U_{\alpha\alpha}(t) = \text{exp} \left[-i 2J \sigma_{\alpha}^1 \sigma_{\alpha}^2  t \right] $. The $U_{YY}(t)$ evolution can be implemented along the same lines. \\
By combining $U_{XX}(t)$ or $U_{YY}(t)$ with a proper rotation of one of the two qubits, a generic spin-spin interaction of the form $\sigma_\alpha^1\sigma_\beta^2$ (with $\alpha\neq\beta$) can be obtained. For instance 
\begin{equation}
R_y^{2}\left(-\frac{\pi}{2}\right)   U_{XX}(t) R_y^{2}\left(\frac{\pi}{2}\right)  = e^{-i2J\sigma_x^1\sigma_z^2}.
\end{equation}
Conversely, introducing $R_y^{12}(\pi/2)$ rotations results in the operator $U_{ZZ}(t)$ :
\begin{equation}
U_{ZZ}(t) = R_x^{12}\left(\frac{\pi}{2}\right) U_{XX}(t) R_x^{12}\left(-\frac{\pi}{2}\right)  .
\end{equation}


\begin{thebibliography}{99}
\bibitem{random}
In all the simulations, the system is initialized in a random superposition of all the 4 states in the computational basis.
\end{thebibliography}
\end{document}